\pdfoutput=1
\documentclass[bibyear]{aa}  
\usepackage{graphicx}
\usepackage{txfonts}
\begin{document} 

\authorrunning{C. E. Alissandrakis etal}
\title{A tiny event producing an interplanetary type III burst}
\author{C. E. Alissandrakis\inst{1}, A. Nindos\inst{1}, S. Patsourakos\inst{1},
A. Kontogeorgos\inst{2} \and P. Tsitsipis\inst{2}
}

\institute{Department of Physics, University of Ioannina, GR-45110 Ioannina, 
Greece\\
\email{calissan@cc.uoi.gr, anindos@uoi.gr, spatsour@cc.uoi.gr}
\and Technological Educational Institute of Lamia, 35100 Lamia\\
\email{akontog@mail.teiste.gr, tsitsipis@teilam.gr}
}

\date{Received ...; accepted ...}

 
  \abstract
   {}
{We investigate the conditions under which small scale energy release events in the low corona gave rise to strong interplanetary (IP) type III bursts.}
{We analyze observations of three tiny events, detected by the Nan\c cay Radio Heliograph (NRH), two of which produced IP type IIIs. We took advantage of the NRH positioning information and of the high cadence of AIA/SDO data to identify the associated EUV emissions. We measured positions and time profiles of the metric and EUV sources.} 
{We found that the EUV events that produced IP type IIIs were located near a coronal hole boundary, while the one that did not was located in a closed magnetic field region. In all three cases tiny flaring loops were involved, without any associated mass eruption. In the best observed case the radio emission at the highest frequency (435\,MHz) was displaced by $\sim$55\arcsec\ with respect to the small flaring loop. The metric type III emission shows a complex structure in space and in time, indicative of multiple electron beams, despite the low intensity of the events. From the combined analysis of dynamic spectra and NRH images we derived the electron beam velocity as well as the height, ambient plasma temperature and density at the level of formation of the 160\,MHz emission. From the analysis of the differential emission measure derived from the AIA images we found that the first evidence of energy release was at the footpoints and this was followed by the development of flaring loops and subsequent cooling.}
{Even small energy release events can accelerate enough electrons to give rise to powerful IP type III bursts. The proximity of the electron acceleration site to open magnetic field lines facilitates the escape of the electrons into the interplanetary space. 
The offset between the site of energy release and the metric type III location warrants further investigation.}

   \keywords{Sun: radio radiation -- Sun: UV radiation -- Sun: flares -- Sun: corona -- Sun: magnetic fields}

   \maketitle
%

\section{Introduction}

The reconfiguration of solar magnetic field to a lower energy state,
which may be facilitated by magnetic reconnection, is the source of
large amounts of energy released during solar flares. This energy
leads to particle acceleration, either in the energy release region
itself, or its vicinity.  Beams of energetic electrons can form either
from the acceleration mechanism  itself or through the propagation
process. Instabilities in the beam may  generate Langmuir waves that
can be transformed in part into electromagnetic waves at the plasma
frequency or its second harmonic via scattering off thermal ions or,
more likely, coupling to low-frequency waves. This is the plasma
emission mechanism; for more details, see the books by Melrose
(1980); Tsytovich (1995) and reviews by Dulk (1985); Bastian et
al. (1998); Nindos et al. (2008); Melrose (2009); Reid \& Ratcliffe
(2014).  

In radio dynamic spectra, the plasma emission from beams of energetic
electrons usually appears as intense bands of emission drifting
rapidly to low or high frequencies depending on whether the beams
travel away or toward the Sun.  Bursts drifting to lower frequencies
are called type III bursts whereas those drifting to higher
frequencies are called reverse-slope bursts.  Type III bursts
propagate out to regions of very low density and therefore the exciter
corresponds to radio-emitting beams that escape upwards along open
magnetic field lines. If the beam of electrons travels along a closed
loop, the plasma density gradient, and thus the frequency drift,
reverses yielding J-bursts of U-bursts (e.g. Labrum \& Stewart
1970). N-bursts (Caroubalos et al. 1987) are also occasionally seen,
when a second reversal of the gradient occurs due to magnetic
mirroring in the loop footpoints.

Type III bursts have been observed from frequencies as high as $\sim$1
GHz at the base of the corona to about 30 kHz at 1 AU and even lower
further out. In several cases type III bursts from the corona continue
into the interplanetary (IP) medium; these events are indicative of
open field lines emanating from the vicinity of the acceleration
region and extending into the IP medium. However, not every coronal
type III burst, even if strong, produces  an IP type III burst
(e.g. Gopalswamy et al. 1998). This may be due to the  complete energy
loss of the electrons. 

The frequency drift rate of type III bursts can be directly converted
into a drift of ambient coronal density with height, and the electron
beam speed can be computed if a coronal density model is provided. The
speeds derived range from about $c/3$ in the low corona to about
$c/10$ at 1 AU.  

The properties of type III bursts have been used for the diagnosis of
the electron acceleration region. Aschwanden \& Benz (1997) inferred
the ambient coronal density of the acceleration region from the plasma
frequency at the separatrix between correlated type III bursts and reverse-slope
bursts. Their results indicated that electrons are accelerated above
the soft X-ray flaring loop-tops. More recently, this result was
refined by Reid et al. (2011; 2014) who exploited the connection
between type III starting frequency and HXR spectral index and found
that the  height of the acceleration region varied between 25 to 200
Mm.

Type III bursts usually come from active regions (e.g. Saint-Hilaire
et  al. 2013) and usually appear in groups of some ten or more
bursts. A large fraction of flares, particularly impulsive flares
(e.g. Cane \& Reames 1988), exhibit type III bursts at their onset.
There are several publications about the statistical relationship
between type III bursts and HXR bursts (e.g. Kane 1972; 1981; Hamilton
et al. 1990; Aschwanden et al. 1995). More recently, Benz et al.
(2005; 2007) studied a large number of flares above GOES class C5, and
found that nearly all flares were associated with coherent radio
emission between 4 GHz to 100 MHz; most of their flares with coherent
radio emission had type III bursts. 

\begin{figure}
\centering
\includegraphics[width=\hsize]{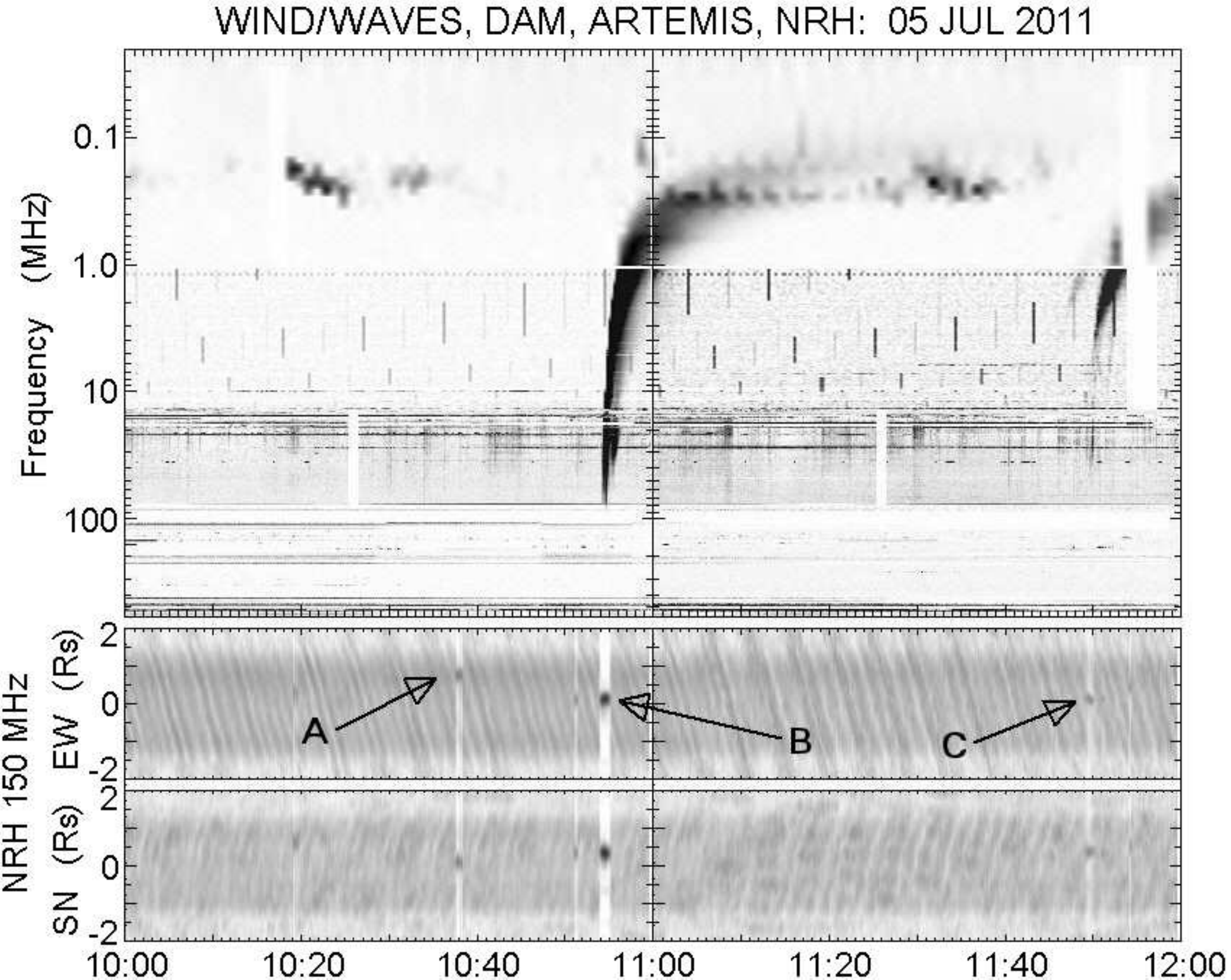}
\caption{Composite Dynamic spectrum and NRH 1D images 
(from the {\it Radio Monitoring} site, http://secchirh.obspm.fr/index.php)
}
\label{comp}
\end{figure}

Type III radio bursts are known to be associated not only with flares,
but also  with H$\alpha$ ejecta (Axisa et al. 1973), X-ray bright
points (Kundu et al.  1994; 1995a), soft X-ray transient brightenings
(Nindos et al. 1999), and  soft X-ray or/and EUV jets (Kundu et
al. 1995b;  Raulin et al. 1996; Pick et al. 2006; Nitta et al. 2008;
Innes et al. 2011;  Krucker et al. 2011; Klassen et al. 2011; 2012;
Chen et al. 2013a,b). Jets that are correlated with type III bursts
accompany flaring activity that ranges from weak transient brightenings and
subflares (Kundu et al. 1995b; Raulin et al. 1996; Nitta et al. 2008)
to  M-class flares (Krucker et al. 2011; Chen et al. 2013a). Their
dimensions may range from about 10$^4$ km to more than about 10$^5$
km. Usually they  appear at the edge of active regions and sometimes
(Pick et al. 2006) close to the border of an active region and a coronal
hole. 

The correlation of type III bursts with soft X-ray/EUV jets and weak
transient brightenings indicates that particle acceleration may take
place when jets and weak transient brightenings occur. This is also
supported by the in-situ detection of electron spikes at energies
below 300 keV (Klassen et al. 2011; 2012) whose timing is well
correlated with type III bursts and EUV jets. For jet events
correlated with both type III bursts and electron events observed in
situ, the flare geometry can be explained by reconnection between open
and closed magnetic field lines (the so-called interchange
reconnection; Krucker et al. 2011).  The previously open field lines
form the flaring loops, whereas heating along the newly opened field
lines is weaker because material can easily be lost due to the open
geometry of the field. Type-III-producing electrons escape along the
newly opened field line. 

In this article we analyze three tiny events detected by the
Nan\c{c}ay Radio Heliograph (NRH). Two of the events were associated
with type III bursts that could be traced from the corona into the IP
medium. These events are remarkable because they show that strong IP
type III bursts may originate in electrons released in very weak
events, provided that the electrons get access to open field
lines. The article is organized as follows: in section 2 we describe the observations and present the results,
while in section 3 we discuss the conclusions.

\section{Observations and results}
Figure \ref{comp} gives a composite dynamic spectrum in the range 700 MHz to 0.02 kHz, together with 1-dimensional images (EW \& NS) from the NRH. In the time interval of 2 hours shown in the figure, three weak emissions are seen in the NRH 1D images; out of these, events B and C were accompanied by interplanetary type IIIs, while A was not. Note that, apart from these three weak emissions, the sun was very quiet at metric wavelengths, with the quiet sun clearly visible in the NRH 1D images.

\subsection{EUV and x-ray data for event B}
\begin{figure*}
\centering
\includegraphics[width=\hsize]{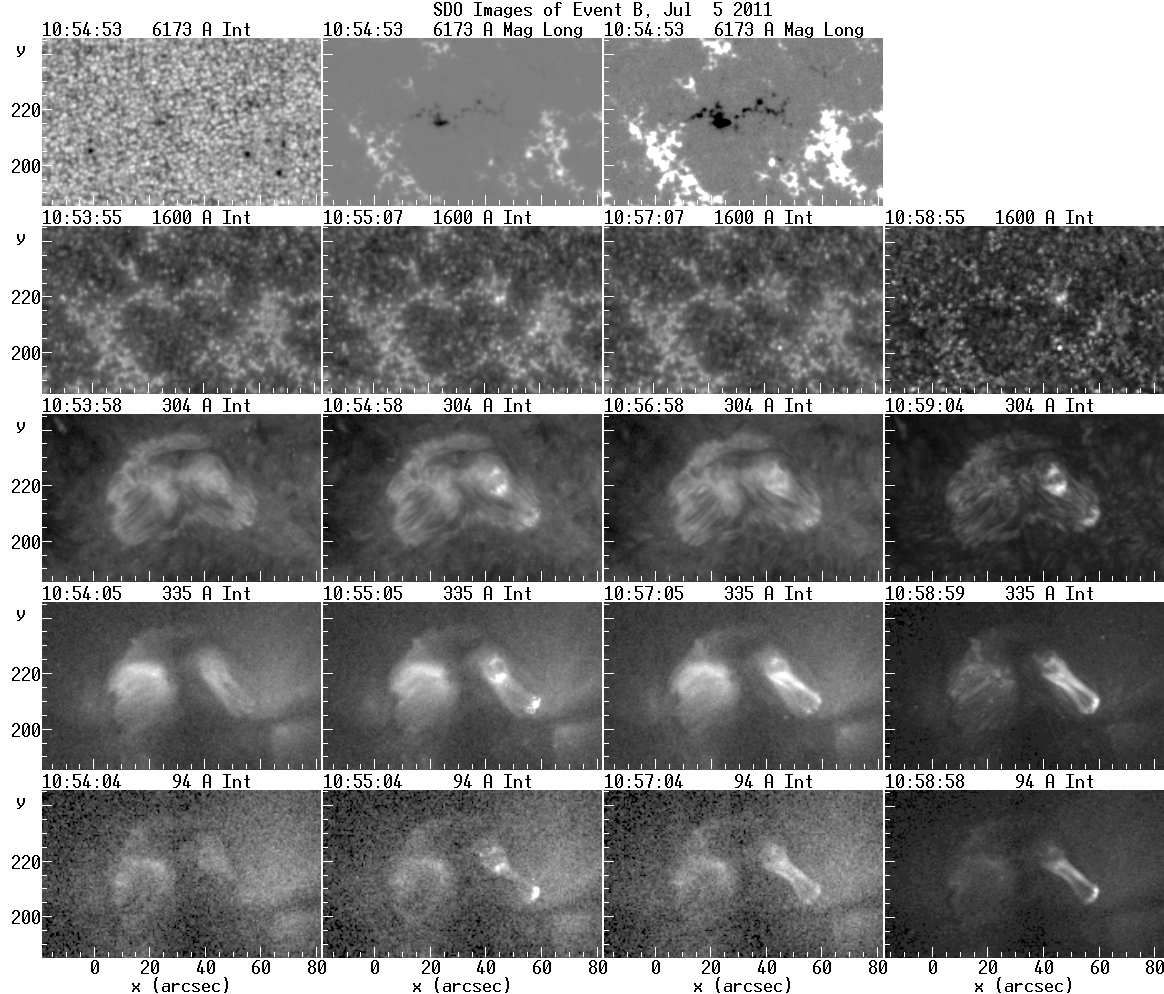}
\caption{SDO images of the flaring loop associated with event B. The top row shows an HMI white light image with the corresponding magnetogram shown both with the full intensity range ($\pm975$\,G) and saturated at $\pm200$\,G. The other rows show AIA images in the 1600, 304, 335 and 94\,\AA\ bands. The right-most column gives the RMS of the intensity during the event at the corresponding band. In this and all figures the contrast for each wavelength band has been adjusted for best viewing.}
\label{sdo1}
\end{figure*}

Based on the NRH positions, we identified in SDO/AIA images a small flaring loop, $\sim20$\arcsec\ long, located at 3.2\,W, 16.4\,N, as the source of the strongest event, B (Fig. \ref{sdo1}). The HMI continuum image (top row) shows a group of small pores and the magnetogram shows two negative polarity patches surrounded by positive polarity magnetic field. Before the event (left column) there are two sets of loops, joining each negative patch with the nearby positive polarities. 

There was no detectable emission in the 1700\,\AA\ band. In all other AIA wavelength bands the event started with brightenings at the footpoints located in opposite polarity regions within the west set of pre-existing loops; this was followed by emission along the entire loop in all bands except 1600 and 304\,\AA. The overall structure of the emission is well visible in the intensity RMS images, shown in the right-most column of Figure \ref{sdo1}; here we can see that in the high temperature channels the loop consists of at least two distinct strands, which are not parallel to each other but their projected distance varies from $\sim3.5$\arcsec\ near the middle of the loop to $\sim5.5$\arcsec\ at the footpoints. The footpoints themselves consist of at least two components.

\begin{figure}
\includegraphics[width=4.5cm]{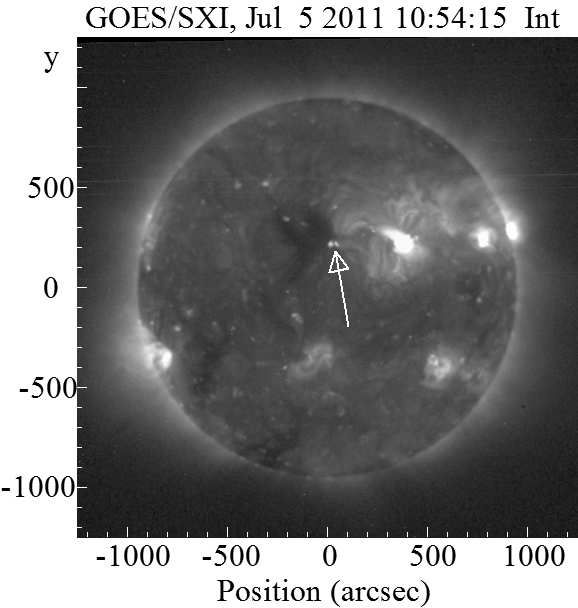}

\vspace{-6.5cm}\hspace{4.3cm}
\includegraphics[width=4.4cm,bb=0 0 400 500,angle=-90]{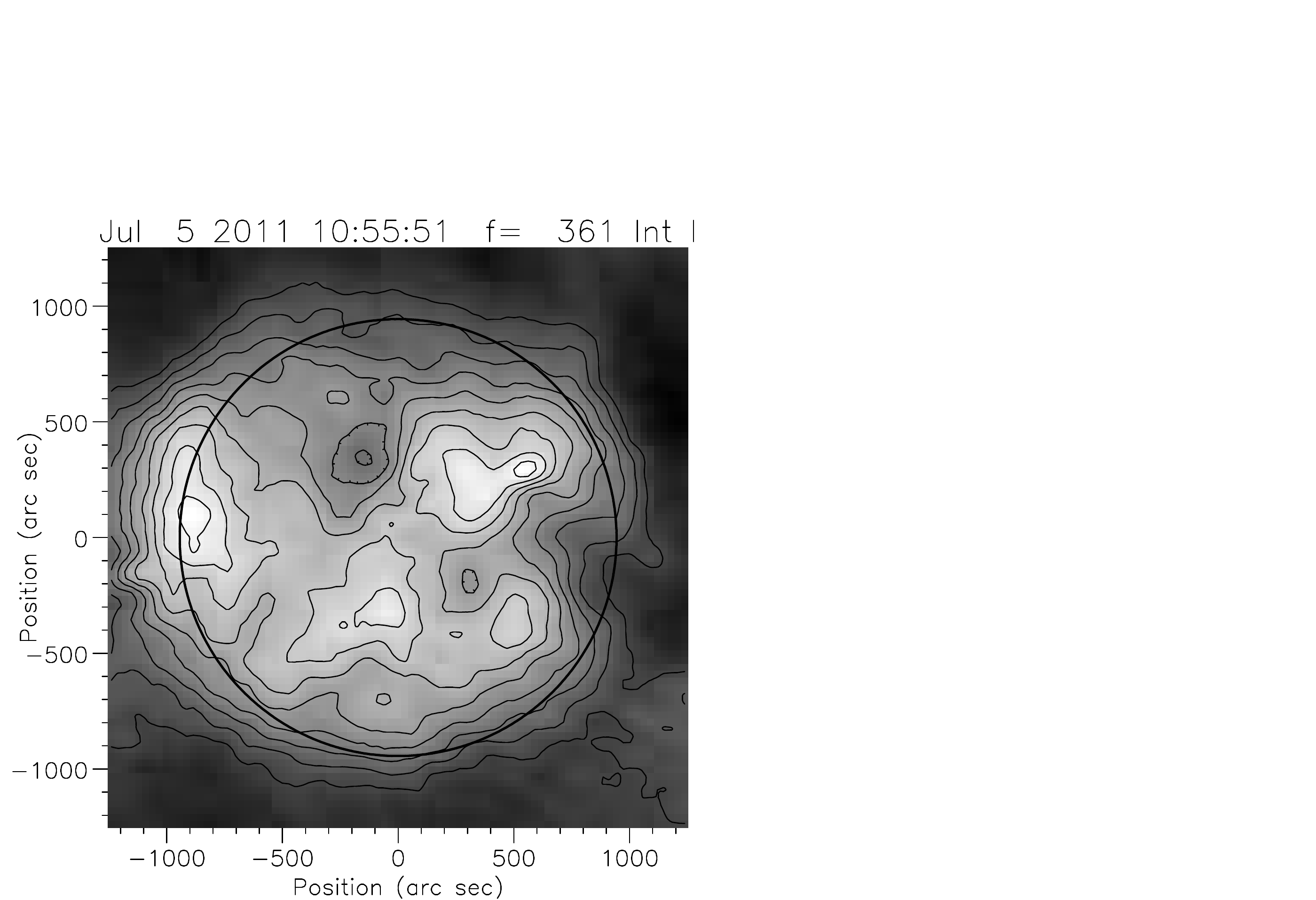}

\vspace{2cm}
\caption{Full disk images of the quiet sun: SXI, TM filter (left) and NRH, 361\,MHz (right, contours from 10$^5$ to $9\times10^5$\,K in steps of 10$^5$\,K). The arrow in the SXI image points to the two pre-flare bright points }
\label{SXI-NRH}
\end{figure}

No emission associated with the event was detected in the GOES flux data in the 1 to 8\,\AA\ band above the noise level of $4\times10^{-9}$\,Wm$^{-2}$, while the total flux of the sun in the 0.5 to 4\,\AA\ band was below the  detection limit of $10^{-9}$\,Wm$^{-2}$. We also found no trace of the event in the RSTN microwave patrol data. RHESSI was in the earth's shadow at the time, thus we have no hard x-ray information. Hinode was not observing the region at the time of the events, while GOES/SXI images show the two sets of loops mentioned above as bright points and our event is barely discernible as a very weak brightening of the western bright point. The SXI images show that the flaring loop, located near the central meridian, was close to the border of a small coronal hole which was also visible in the NRH quiet sun image (Fig. \ref{SXI-NRH}).

\begin{figure}
\centering
\includegraphics[angle=-90,width=\hsize]{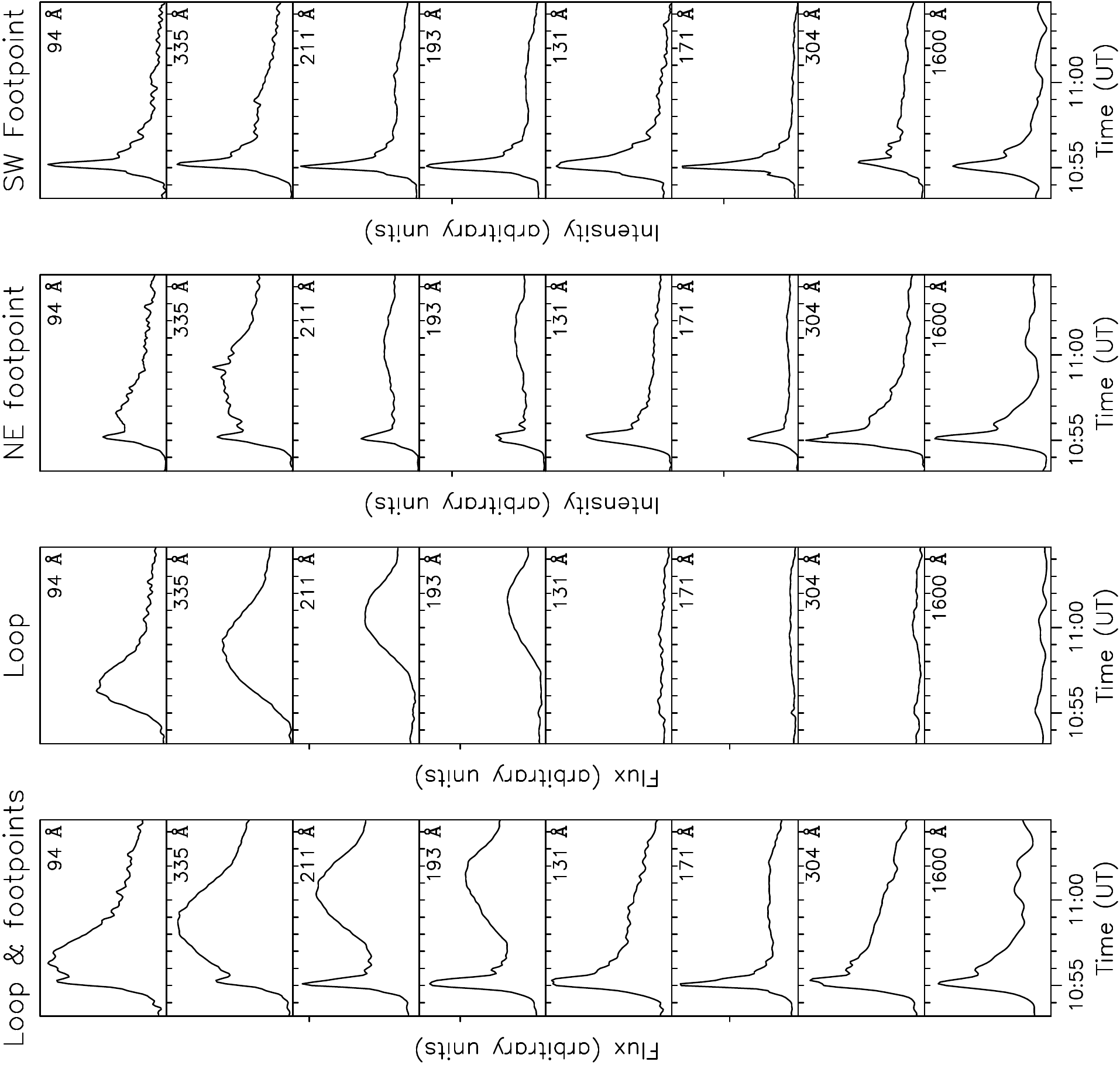}
\caption{Flux of the entire event B, the loop and the footpoints as a function of time for the AIA wavebands}
\label{prof1}
\end{figure}

Time profiles of the event in the 1600\,\AA\ and the EUV bands of AIA, integrated in the direction perpendicular to the loop, are presented in Figure \ref{prof1}. Their most important aspect is that the flux profiles of the entire event (left column in the figure)  show, in addition to an impulsive component, a ``post burst'' increase,  most prominent in the 193, 211, 335 and 94\,\AA\ channels. 

As evidenced by the time profiles of the loop (second column in Fig. \ref{prof1}), the post burst is associated with the body of the loop, while the  time profiles of the footpoints (third and fourth column in the figure) were impulsive; there is some post-burst emission from the NE footpoint, but this is, most probably, due to the fact that this foopoint cannot be fully separated from the loop body due to projection effects. The emission from the body of the loop appeared first in the hot AIA band of 94\,\AA\ and gradually in the cooler 335, 211 and 193 bands. This behavior is similar to that of two weak events described by \cite{2013A&A...556A..79A}, which had one footpoint inside a sunspot umbra and loop lengths of 44 to 53\arcsec; they were interpreted in terms of coronal loop cooling, following an impulsive heating to multi-million K temperatures, with the heating phase being unobservable due to the low emission measure associated with it. 

\begin{figure*}
\sidecaption
\centering
\includegraphics[width=12cm]{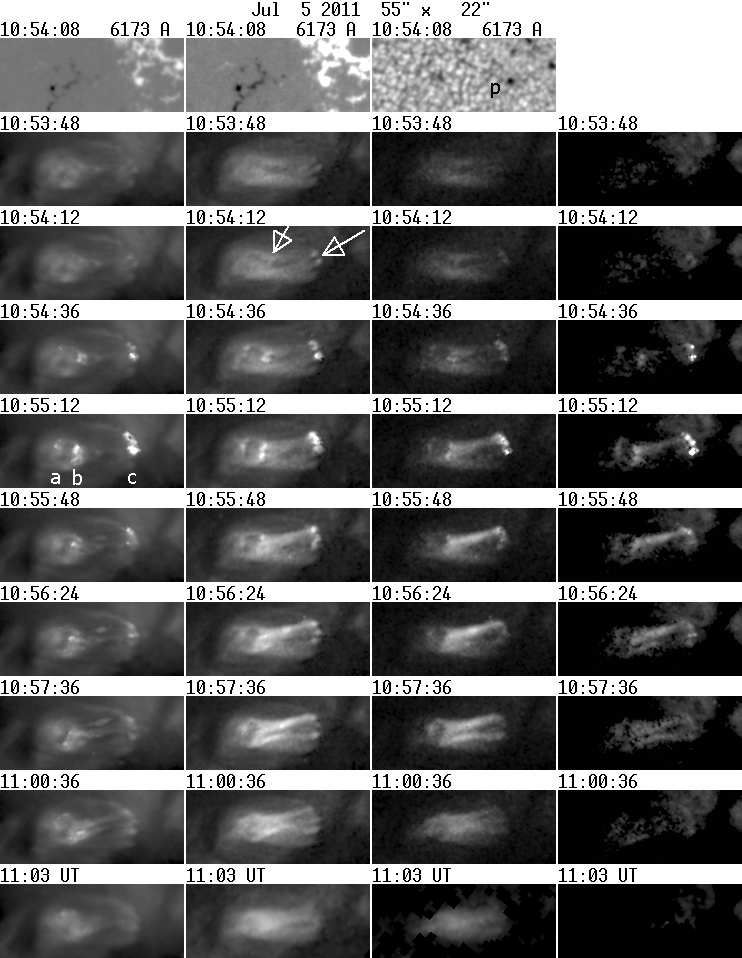}
\caption{Selected images of the DEM in the temperature ranges 1-2\,10$^6$\,K (first column, maximum value $2.5\times 10^{28}$\,cm$^{-5}$), 2-4\,10$^6$\,K (second column, maximum value $3.0\times 10^{28}$\,cm$^{-5}$), 4-8\,10$^6$\,K (third column, maximum value $2.0\times 10^{28}$\,cm$^{-5}$) and above 8\,10$^6$\,K (right column, maximum value $2.0\times 10^{28}$\,cm$^{-5}$) in the course of event B. The contrast of the display (0.6 for the first column and 0.4 for the others) was adjusted for best viewing. The top row shows HMI magnetograms (full range and saturated at $\pm300$\,G) and a while light image. The arrows in the 10:54:12 UT image mark the first brightenings, {\bf a}, {\bf b} and {\bf c} in the 10:55:12 UT image mark three sets of footpoints, {\bf p} in the white light image marks a small pore. The field of view is 55 by 22\arcsec\ and the images have been rotated for easier viewing. A movie available in the online edition shows the full time sequence of DEM images (top row of movie) as well as DEM difference images (bottom row of the movie).
}
\label{DEM}
\end{figure*}

Although a detailed analysis of the event in the EUV is beyond the scope of this article, we would like to report here some interesting results derived from a differential emission measure (DEM) analysis of the AIA images, which we performed using the algorithm of  \cite{2013ApJ...771....2P}. This is the same algorithm used by \cite{2013A&A...556A..79A} and by \cite{2013PASJ...65S...8A}; we obtained a satisfactory fit to the data for most of the field of view, with somewhat less satisfactory fit at the footpoints. Figure \ref{DEM} shows selected DEM images in four temperature ranges (1-2$\times10^6$\,K, 2-4$\times10^6$\,K, 4-8$\times10^6$\,K and above 8$\times10^6$\,K), together with HMI magnetic field and intensity images. 
A time sequence of DEM images and differences is shown in the movie attached to Fig. 5.

Before the event a system of loops was visible at temperatures between 2 and 4$\times10^6$\,K, spanning opposite magnetic polarities, with the rightmost end of the loops being near a small pore. The first brightenings appeared at 10:54:12 UT at locations {\bf b} and {\bf c} (arrows in Fig. \ref{DEM}). Note that, although location {\bf c} was near the rightmost edge of the pre-existing loops, location {\bf b} was not near the leftmost edge (location {\bf a} in fig. \ref{DEM}), but was rather associated with a weak negative polarity patch in between; brightenings at location {\bf a} appeared $\sim24$\,s later, at 10:54:36 UT. 

Loop emission first appeared at T$>8\times10^6$ near the upper footpoint at location {\bf b} and quickly spread to the upper footpoint at location {\bf c} with a projected velocity (hence a lower limit to the true velocity) of $\sim400$\,km\,s$^{-1}$, becoming stronger at lower temperatures while withering away at high temperatures. A second loop appeared around 10:56:26 near the lower footpoint at location {\bf b} and spread towards the lower footpoint at location {\bf c} with an apparent velocity of $\sim80$\,km\,s$^{-1}$ in the $4\times10^6$\,K$<T<8\times10^6$\,K range. Later in the event the emission of the upper loop extended to footpoint {\bf a}. Note that the footpoint emission had weakened considerably by the time the flaring loops had fully developed. Note also that both flaring loops were inclined by about 10\degr\ with respect to the pre-flare loops. 

The overall picture is that the first evidence of energy release was the heating of the footpoints to temperatures of $\sim10^7$\,K;  this was followed by the loop emission, first appearing at high temperatures and cooling gradually. The apparent expansion of the flaring loops from footpoint {\bf b} to footpoint {\bf c} is suggestive of evaporation of material heated in the course of the energy release process ({\it c.f.} \cite{1984ApJ...281L..79F}).

\subsection{Radio observations}\label{radio_obs}
\begin{figure}
\centering
\includegraphics[width=\hsize]{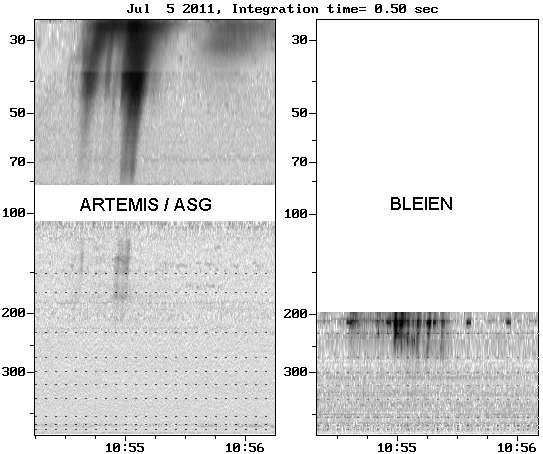}
\caption{Dynamic Spectra from ARTEMIS-IV and Bleien. Dotted lines mark the frequencies of the NRH. The Bleien data are  from the Callisto site http://soleil.i4ds.ch/ solarradio/callistoQuicklooks/.}
\label{dyn}
\end{figure}

Going back to the metric data, we show in Figure \ref{dyn} dynamic spectra of the event in the metric-decametric range from the ARTEMIS sweep frequency receiver (ASG) and from the Bleien spectrograph. Unfortunately, the high sensitivity, high cadence acoustico-optic receiver of ARTEMIS (SAO) was not operating. 

\begin{figure}
\includegraphics[angle=-90,width=5.6cm]{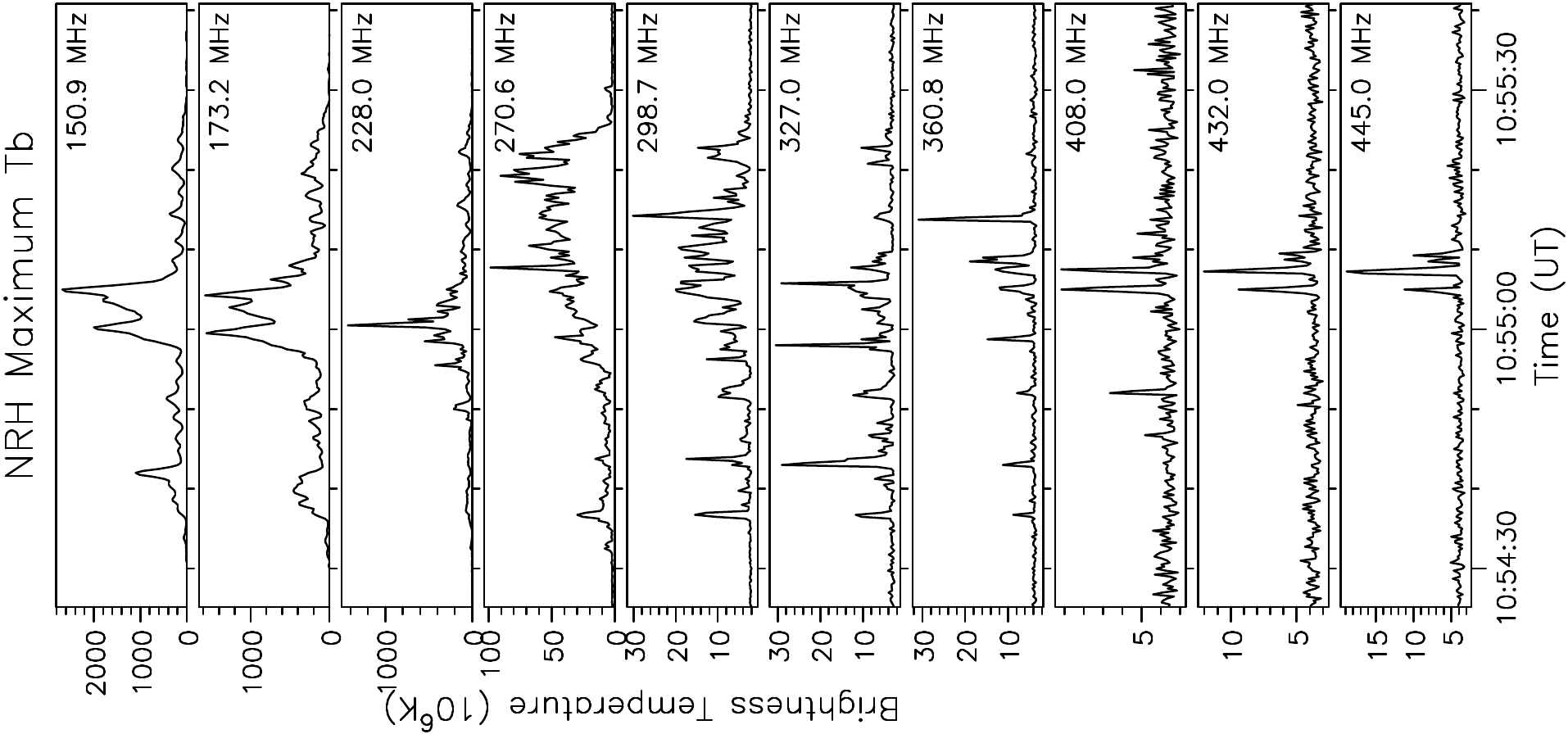}%

\vspace{-7.07cm}\hspace{5.7cm}\includegraphics[angle=-90,width=3cm]{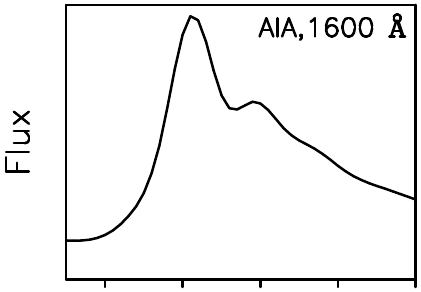}

\hspace{5.7cm}\includegraphics[angle=-90,width=3cm]{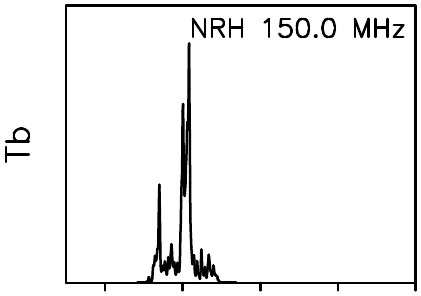}

\hspace{5.7cm}\includegraphics[angle=-90,width=3.24cm]{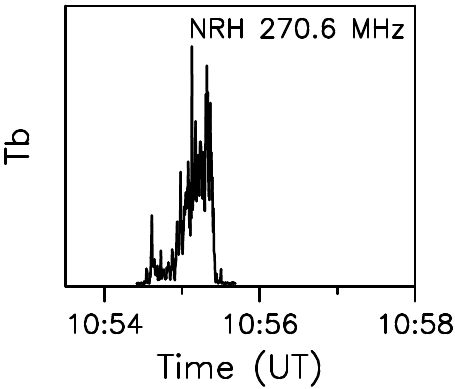}
\caption{Maximum brightness temperature as a function of time; note that the scales are different for each frequency (left). Comparison of the 270.6 and 150 light curves with that of the 1600\,\AA\ AIA band (right).}
\label{profs}
\end{figure}

In the decametric range the dynamic spectrum shows two groups of Type IIIs, followed by a Type V. In the metric range the type III emission is weak and is detectable up to 200 MHz in the ARTEMIS spectrum and up to about  300 MHz in the Bleien spectrum.  The principal Type III group shows a fairly complex structure with several components, apparently corresponding to individual electron beams. This complex structure is more apparent in the much more sensitive  NRH data, and the brightness temperature as a function of time is shown in Figure \ref{profs}, left. The maximum is at 150.9\,MHz  ($2.7\times10^9$\,K) and the weakest peak is at 408\,MHz ($9.4\times10^6$\,K). Note the great decrease of brightness, by  a factor of about 300, between low and high frequencies; in particular, there is a brightness drop by more than a factor of 10 between 228 and 270.8\,MHz that makes the time profiles look different. Note also that the maximum brightness temperature of type III bursts is reported to range from 10 MK up to about 10$^{12}$ K for coronal type III bursts (e.g. Saint-Hilaire et al. 2013) and up to about 10$^{15}$ K in interplanetary type III bursts (e.g. Melrose 1989).

Compared to the EUV evolution (Fig. \ref{profs}, right), the principal beams appeared during the rise phase of the impulsive component of the EUV emission, with some other beams, seen best at 270.7 and 327 MHz, extending slightly over the maximum.

\begin{figure}[h!]
\centering
\includegraphics[width=\hsize]{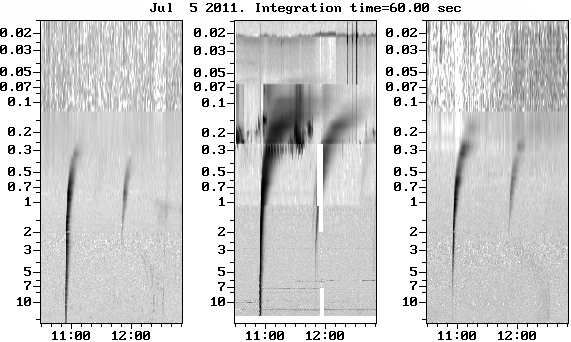}
\caption{Dynamic spectra of events B and C from SWAVES-A (left), WIND/WAVES (middle) and SWAVES-B (right). The frequency is in MHz.}
\label{waves}
\end{figure}

The dynamic spectra of events B and C, as recorded by the WAVES instruments on STEREO A, WIND and STEREO B are given in Figure \ref{waves}. WIND/WAVES (\cite{1995SSRv...71..231B}) has a thermal noise receiver (TNR), operating in the 4-256\,kHz range and two radio band receivers, RAD1 operating from 20 to 1040\,kHz and RAD2 from 1.075 to 13.825\,MHz; the low frequency receiver of STEREO/WAVES (\cite{2008SSRv..136..487B}) (LFR) operates in the 2.5 to 160\,kHz range and the two high frequency receivers operate from 125 to 1975\,kHz (HFR1) and from 2025 kHz to 16.025\,MHz. At the time of our events STEREO A was 97.6\degr\ ahead of the earth and STEREO B 92.5\degr\ behind.

Both bursts are visible in all three dynamic spectra. Event B is detectable down to about 100\,kHz in the WIND spectrum. 
Assuming that the electron density is inversely proportional to the square of the distance, which is valid for large distances (\cite{1998SoPh..183..165L}) and under conditions of constant solar wind speed, the frequency is related to the distance from the sun, $r$, as:
\begin{equation}
r=r_E s f_{p,E} f^{-1}
\end{equation}
where $r_E$ is the distance of the Earth, $s$ is 1 for the emission at the plasma frequency and 2 for harmonic emission, $f$ is the frequency of observation and $f_{p,E}$ is the plasma frequency of the solar wind near the Earth. The latter can be measured from the TNR data (horizontal ridge in Fig. \ref{waves}) and its value was 21\,kHz in our case. Hence the 100 kHz level is at $r\sim45$\,R$_\odot$ for $s=1$ ($n_e=180$\,cm$^{-3}$) and $\sim90$\,R$_\odot$ for harmonic emission ($n_e=45$\,cm$^{-3}$) In SWAVES-A the burst is detectable down to $\sim250$\,kHz and in SWAVES-B down to $\sim120$\,kHz, the differences between them and with WIND/WAVES being due to directivity and/or sensitivity effects.

We computed the average beam velocity from the variation of the time of maximum emission as a function of frequency. We assumed that the electrons moved along a Parker spiral, rooted near the flaring loop and that the magnetic field became radial at 3\,R$_\odot$; we used Eq. (1) to associate the frequency with the distance from the sun, we adopted a solar wind speed of 450\,km\,s$^{-1}$ and took into account the light travel time. Refraction effects were ignored in this computation; these would be stronger for emission in the fundamental.

The fit was better for emission at the harmonic of the plasma frequency and, combining the data from all three satellites, we obtained a value of 41\,Mm\,s$^{-1}$ for the beam velocity along the magnetic field lines of force at 3\,R$_\odot$ and a deceleration of -15\,km\,s$^{-2}$, with estimated uncertainty of 1\,Mm\,s$^{-1}$ and 1\,km\,s$^{-2}$ respectively; thus the velocity dropped to 39\,Mm\,s$^{-1}$ at 10\,R$_\odot$ and to 25\,Mm\,s$^{-1}$ at 50\,R$_\odot$. Fitting the data of each individual satellite, we obtained a velocity range of 39-41\,Mm\,s$^{-1}$ and deceleration in the range of -16 to -27\,km\,s$^{-2}$. If, instead, the emission was at the fundamental, the corresponding values of the velocity and the deceleration are 19\,Mm\,s$^{-1}$ and -7\,km\,s$^{-2}$.

Thus the electron beam associated with our event had a moderate speed, $\sim0.14$c. This result is consistent with previous studies: derived speeds for type III bursts can range from $> 0.5c$ in the corona (Poqu\'{e}russe 1994; Klassen et al. 2003) down to about $0.1c$ near the Earth (e.g. Dulk et al. 1987).  It agrees very well with two particular statistical studies of the electron beam speeds inferred from decimetric type III bursts (Aschwanden et al. 1995; Mel\'{e}ndez et al. 1999).

As far as the flux of the emission is concerned, we made an estimate at 1\,MHz by comparing our data with those of two bursts shown in figures 3 and 4 of \cite{2014SoPh..289.3121K}. We obtained flux values of about $4\times10^{-18}$\,Wm$^{-2}$Hz$^{-1}$ and $0.9\times10^{-18}$\,Wm$^{-2}$Hz$^{-1}$ for SWAVES A and B respectively; compared to the flux distribution derived from 152 bursts by \cite{2014SoPh..289.3121K}, our event is near the average, which at 1\,MHz is about  $2.5\times10^{-18}$\,Wm$^{-2}$Hz$^{-1}$.

\subsection{Burst positions in the metric range}
\begin{figure}
\centering
\includegraphics[width=6cm]{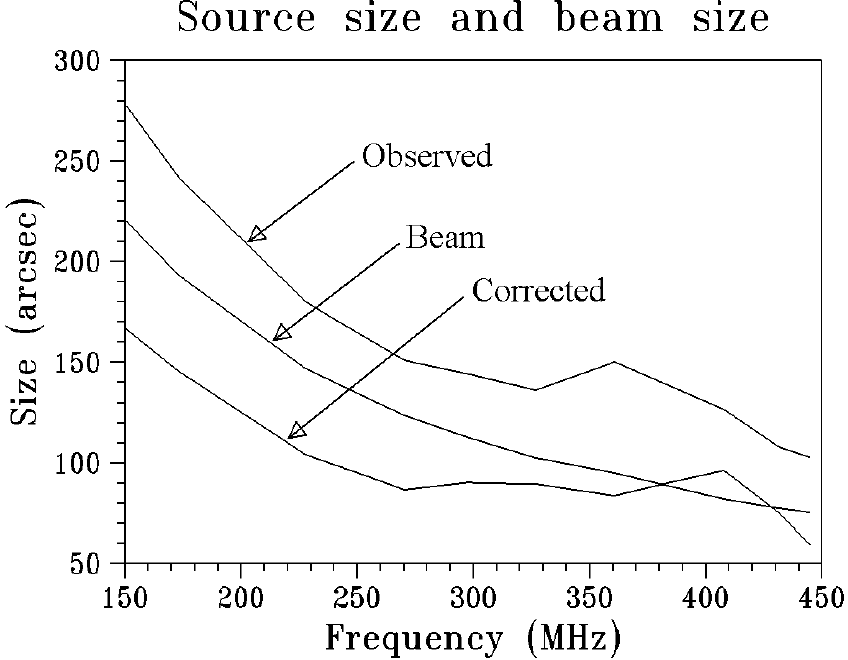}
\caption{Observed source size, beam size and corrected source size.
}
\label{size}
\end{figure}

We computed 2D maps from the NRH visibilities. For this, it was necessary to introduce phase corrections to the visibilities, which were derived by self-calibration, using strong point-like sources. In all frequencies the sources had a simple structure; they were slightly resolved and their size decreased with frequency (Fig. \ref{size}).

\begin{figure}[h]
\centering
\includegraphics[width=\hsize]{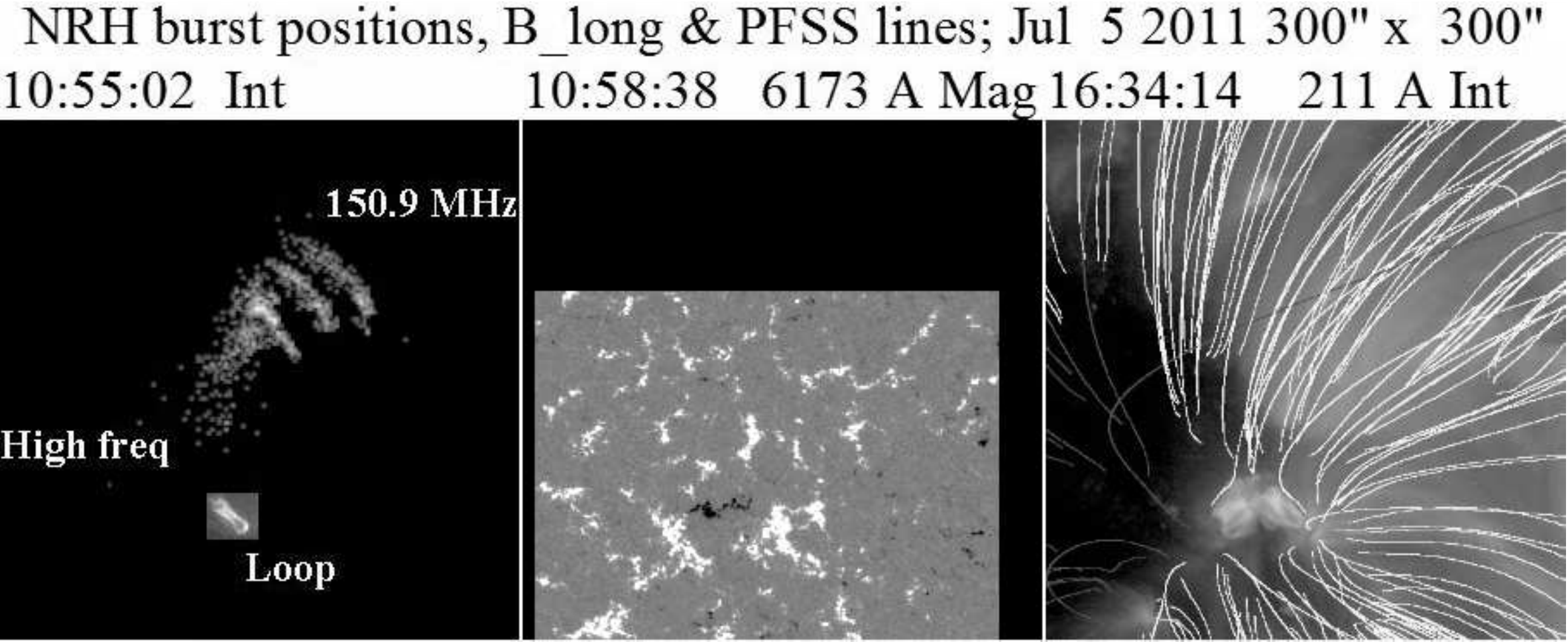}
\caption{From left to right: Combined 2D histograms of the position of NRH sources at all frequencies; the insert shows an image of the loop at 335\,\AA; HMI/SDO magnetogram; lines of force of the potential magnetic field extrapolation superimposed on a 211\,\AA\ AIA image (from http://sdowww. lmsal.com/suntoday/).}
\label{pos}
\end{figure}

From the position of the emission peaks at each frequency we constructed 2-dimensional histograms of the number of peaks as a function of $x$ and $y$ position and put all frequencies together. The result is shown in the left panel of Figure \ref{pos}, together with an image of the loop; also shown in the figure is a magnetogram and an SDO image at 211\,\AA\ with superimposed magnetic field lines. We first note that there is a net drift of positions from high to low frequencies, apparently reflecting the average upward motion of the electron beams along the magnetic field. The fact that we are near the coronal 
hole boundary explains why the beams found open magnetic field lines to escape into the interplanetary space.

The positions at 150.9, 173.2 and 228\,MHz are spread over elongated regions, 10\arcsec\ wide and 70\arcsec\ long; although the extent of the regions is smaller than the instrumental beam (cf. Figure \ref{size}), this displacement is most probably real both because it is not a random spread (it is much longer than it is wide) and, most important, there is a systematic  drift from NW to SE during the 2 min of the radio data. Thus the observed drift apparently reflects the fact that we have a multitude of individual beams, each following a slightly different field line.

Note that there is a net displacement of the radio sources from the flaring loop, which amounts to about 55\arcsec\ for high frequencies. As the radio sources were located in a region of high density gradient, between the coronal hole and the nearby closed magnetic configurations, their observed position may have been affected by refraction. It is outside of the scope of this article to model these effects in detail, still we did some indicative ray tracing computations using simple models described by \cite{1994AdSpR..14...81A}, assuming emission at the harmonic of the plasma frequency. 

We found that, as a result of refraction, the true position of the radio source is closer to the center of the coronal hole than the observed; hence the true distance of the radio sources from the flaring loop is even greater than the observed and this effect will be stronger if the emission is at the fundamental. The computed displacement is $\sim10$\arcsec\ at 400\,MHz and increases to $\sim30$\arcsec\ at 160\,MHz, an effect that will make the true orientation of the magnetic field lines traced by the beam less inclined than suggested in Figure \ref{pos} and closer to the direction of the potential field lines.

\begin{figure*}
\hspace{2.2cm}\includegraphics[angle=-90,width=3.15cm]{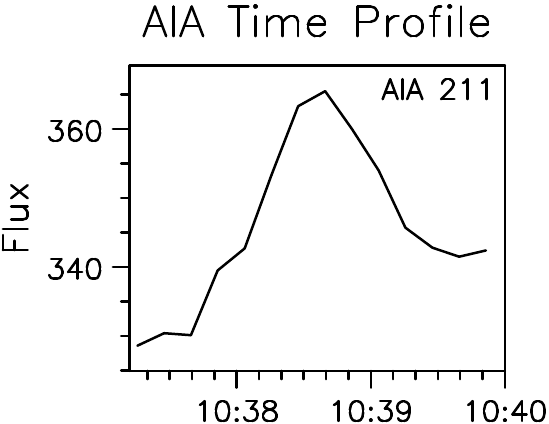}

\hspace{2.2cm}\includegraphics[angle=-90,width=3.15cm]{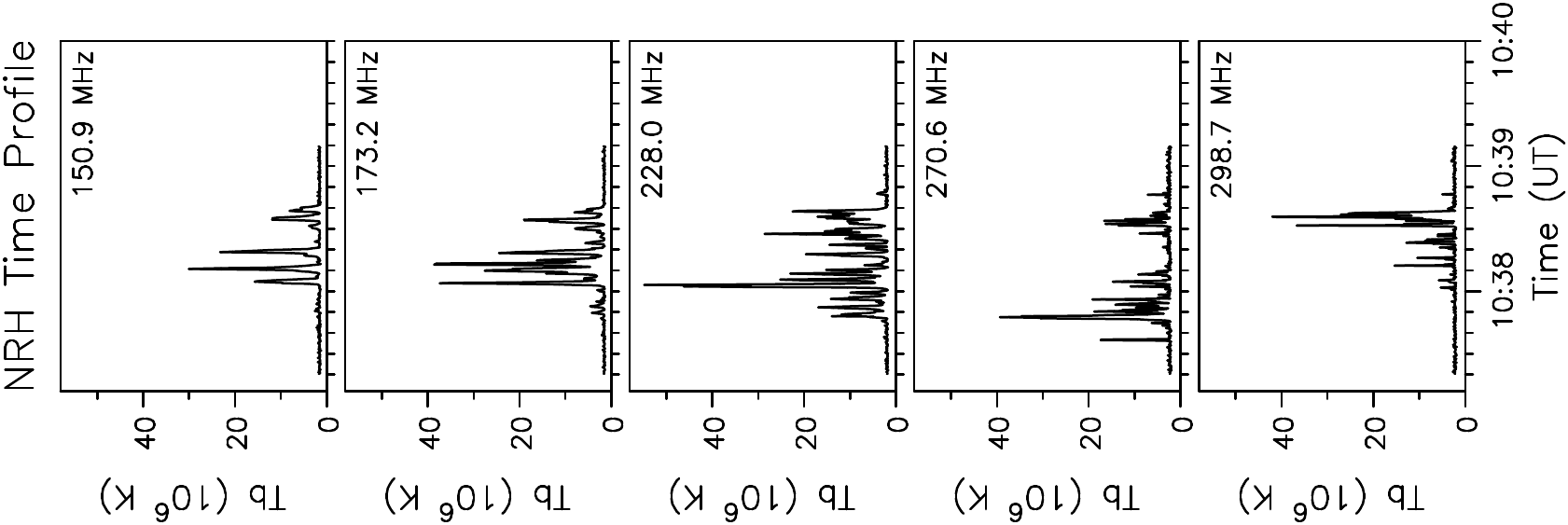}

\vspace{-12.2cm}\hspace{5.5cm}
\includegraphics[width=10.35cm]{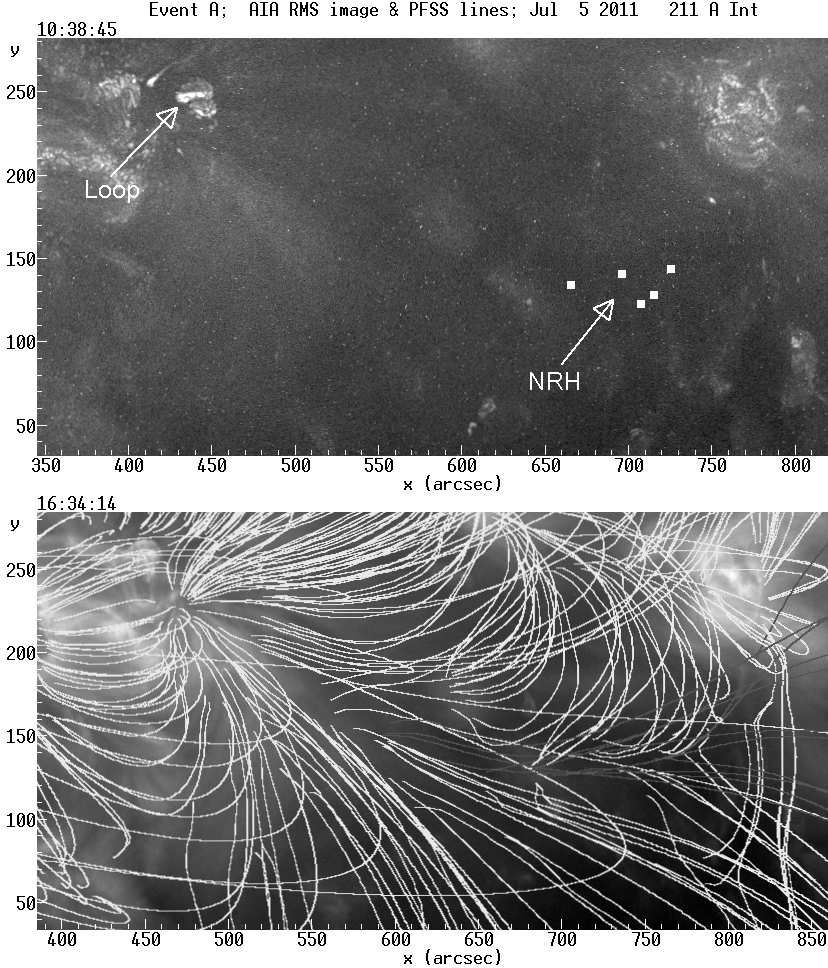}
\caption{Left: AIA and NRH time profiles of Event A. Right, top: Image of the intensity RMS in the 211\,\AA\ band; white 
squares mark the position of the NRH sources at (from E to W) 270.6, 298.7, 150.9, 173.2 and 228\,MHz. Right, bottom: PFSS magnetic lines of force (from http://sdowww. lmsal.com/suntoday/).}
\label{bpa}
\end{figure*}

The velocity of the beam can be estimated from the frequency drift of the type III emission. We obtained consistent values of the drift, $d \ln f/dt=-0.25$\,$s^{-1}$, both from the ARTEMIS spectra and the NRH data around 160\,MHz; noting that
\begin{equation}
v\cos\theta=-2 H \frac{d\ln f}{dt}, 
\end{equation}
where $H$ is the density scale height, $v$ the velocity along the field lines and $\theta$ the inclination of the beam trajectory with respect to the vertical (radial direction). In order to make the observed drift compatible with the velocity of 41\,Mm\,s$^{-1}$  estimated from the WAVES spectra (section \ref{radio_obs}), a scale height of 
\begin{equation}
\frac{H}{\cos\theta}=83 \mbox{~Mm~}=0.119 \mbox{\,R$_\odot$}
\end{equation} 
is required.

Additional information can be obtained from the transverse velocity of the beam, $v_t$, which can be computed from the displacement of the radio images and the delay of the corresponding intensity peaks; ignoring refraction effects, our measurements at 150.9 and 173.2\,MHz, where individual peaks are easy to identify ({\it cf.\/} Fig. \ref{profs}), give
\begin{equation}
v_t=v\sin\alpha=38 \mbox{~Mm\,s$^{-1}$}
\end{equation}
where $\alpha$ is the angle between the beam velocity and the line of sight. From the measured transverse velocity and the value of the beam velocity estimated in section \ref{radio_obs}, we obtain $\alpha=66$\degr\ from the above relation. We note here that the beam velocity deduced from the WAVES spectra for emission at the plasma frequency (19\,Mm\,s$^{-1}$) is incompatible with the measured transverse velocity; hence we may safely conclude that the emission was at the harmonic.

Moreover, the orientation of the projection of the beam trajectory on the plane of the sky can be measured from the position of the type III at high and low frequencies (Fig. \ref{pos}), again ignoring refraction effects. From this and the angle $\alpha$ we can compute the angle between the beam and the radial direction, through a coordinate transform taking into account the heliocentric position of the loop. In this way we obtain $\theta=52$\degr, hence $H=51$\,Mm$=0.073$ \,R$_\odot$ from (3).

Furthermore, the scale height depends on the temperature and the distance from the center of the sun, $r$:
\begin{equation}
H=0.0712T_6\left(\frac{R_\odot}{r}\right)^2 \mbox{~~(in units of $R_\odot$)} 
\end{equation}
where $T_6$ if the temperature in units of $10^6$\,K and a molecular weight $\mu=0.61$ has been assumed. A lower limit of $r$ can be estimated from the projected distance between high and low frequency sources (Fig. \ref{pos}), which is 0.118\,R$_\odot$, and the angles $\alpha$ and $\theta$ computed above. This gives $r\simeq1.0757$\,R$_\odot$, which, substituted into (5), gives an ambient temperature of $1.2\times10^6$\,K at the level of formation of the 160\,MHz emission, that corresponds to an electron density of $3.8\times10^8$cm$^{-3}$ for emission at the harmonic.

The above computations show that we can get self-consistent results from the combined information about the beam velocity from WAVES and ARTEMIS spectra and the NRH images. The derived temperature and emission height are both at the low side of the expected values, but this could be due to the fact that refraction effects could not be taken into account. Still we note that the derived electron density is not much different than that predicted from coronal models: at 1.08\,R$_\odot$ the density is $4.2\times10^8$cm$^{-3}$ according to the Newkirk (1961) model and 1.91$\times10^8$cm$^{-3}$ according to the Saito (1970) equatorial model. Furthermore, the scale height of the Newkirk model corresponds to a coronal temperature of $1.4\times10^6$\,K, while the observed brightness temperature of the quiet sun is $\sim1.\times10^6$\,K (Fig. \ref{SXI-NRH}).

We note that the comparison of the electron beam velocity in the corona and the interplanetary medium was made under the assumption that, in both cases, the electron pitch angle has a narrow distribution along the direction of the magnetic field; however, this may not be valid in the solar wind, due to wave-particle interactions that could broaden the pitch angle distribution and reduce the measured beam velocity.

\subsection{Events A and C}
For the other two events we will restrict ourselves to some general comments without presenting a detailed analysis. Starting with event C, we note that it was associated with a flaring of the same loop that produced event B, hence the similarity of the two events in the WAVES dynamic spectra. This flaring was weaker, which explains the weaker type III emission in the decametric and the hectometric range.

The case of event A is of greater interest; it only appeared in the NRH data (Fig. \ref{comp}) without any counterpart in the dynamic spectra, although it was no less intense than event C which was associated with  a week type III.  There were not any brightenings in the immediate vicinity of the metric emission, but we were able to identify its origin on the basis of the time profiles: it was most probably associated to a small flaring loop located more than 1/3 of a solar radius away (Fig. \ref{bpa}). This loop, which was located about 450\arcsec\ west of the loop that produced events B and C, is well visible in the image of the RMS variation of the intensity in the 211\AA\ AIA band (Fig. \ref{bpa}, top right); note that the jet seen to the east of the loop flared up a few minutes earlier than our event. 

As in the case of Event B, the metric emission consisted of many individual peaks during the rise phase of the EUV light curve  (Fig. \ref{bpa}, left). Its brightness temperature was less than that of Event A and there was no detectable emission beyond 298.7\,MHz. The  metric sources did not show the regular arrangement of Event B, but were rather scattered (Fig.  \ref{bpa}, right top). Without dynamic spectra it is difficult to specify the nature of the emission, still some drifting peaks in the time profiles are indicative of electron beams moving upward. The geometry of magnetic field, presented in the bottom right panel of Figure \ref{bpa}, shows that this loop was embedded in a region of closed magnetic lines. Thus the closed magnetic field configuration explains why the electrons did not find their way into the interplanetary space.

\section{Discussion and conclusions}
Our basic conclusion is that strong type IIIs in the decametric and kilometric range may be produced by electrons released in very weak EUV events. Such events would have been missed in the EUV without the AIA superior cadence and resolution: our strongest event was not detectable in the GOES flux data and was barely detectable as a flaring bright point in GOES/SXI images. In all three cases the  associated events were tiny flaring loops; we did not find any evidence of any mass eruption in terms of a jet, ``micro-CME" or propagating disturbance. 

The principal prerequisite for interpranetary type III emission is that the electron beams find open magnetic field lines to propagate in interplanetary space, as demonstrated by the fact that events A and C that originated near a coronal hole gave interplanetary type IIIs, while event B, located inside a closed magnetic field region, did not. Furthermore, for Event B, we found that the position of the metric sources, even at high frequencies, was displaced by about 55\arcsec\ with respect to the nearest footpoint of the associated flaring loop, the true distance being slightly larger due to refraction effects. A much larger distance of $\sim300$\arcsec\ separated the metric emission of Event A from the associated loop. As the electrons are expected to be accelerated near the energy release site, the obvious question arises of how they got there. 

The relatively large offset between the site of energy release and  the metric type III loci presents a formidable challenge to models. It is well-known that particle transport across the mean magnetic field can be dramatically enhanced (well above excursions of about the electron  gyroradius in a collisional time) in the presence of turbulence because of the braiding  of field lines owing to perpendicular magnetic fluctuations. However, the  displacements we found between the metric sources and the flaring loops appear  too large (55-300\arcsec) to be interpreted in terms of cross-field transport  of the energetic electrons. For example, Kontar et al. (2011) found that the  cross-field transport of electrons with energies 8-25 keV in a flaring loop  yielded excursions that were not larger than 1-2\arcsec. 

The AIA data show no  evidence for any small-scale  CME-like event that could possibly act as driver for the displacement of  the electrons  producing the radio sources.  Furthermore, it is unlikely  that such large displacements were the consequence of systematic errors in NRH positional data, because care was taken to introduce phase corrections to the visibilities, which were derived by self-calibration. It is possible that the energetic electrons were transfered away from the flare site through ultra thin loop bundles (cf. Chen et al. 2013) that were below the AIA resolution and  then gained access to open field lines via interchange reconnection. Clearly, more observations that will combine positional information of type III sources at several frequencies with high spatial and temporal resolution EUV observations are needed in order to check how often and under what circumstances displacements like the one reported here possibly occur.

Even at this low intensity level, the metric emission consists of a multitude of sources, apparently associated with different electron beams, evidenced both from the time structure of the emission and also from the displacement of the source position in the course of the event. Apparently each beam is associated with a different episode of electron acceleration and, possibly, of energy release that all happen during the rise phase of the EUV emission (Fig. \ref{profs}).

Using a simple model of the beam electrons propagating along a Parker spiral in interplanetary space and a density variation proportional to $r^{-2}$, we derived, for Event B, the beam velocity (41\,Mm\,s$^{-1}$) and deceleration (-15\,km\,s$^{-2}$) from the observed time of maximum as a function of frequency in the three WAVES dynamic spectra. Combining this information with the drift rate measured from the metric dynamic spectra and the displacement of the radio sources with frequency and time measured from the NRH images, we obtained self consistent values for the height, the temperature and the density at the level of formation of the 160\,MHz emission. We also found that the beam velocity computed under the assumption of emission at the plasma frequency was smaller than the measured transverse beam velocity, hence we conclude that the emission was at the harmonic.

Last but not least, we obtained interesting results from the computation of the differential emission measure from the AIA images. Our analysis for Event B showed that the energy release occurred near the loop footpoints which were heated to above $10^7$\,K and cooled rapidly; the loop body was formed after the footpoint emission, again with a temperature exceeding $10^7$\,K, and cooled slowly. 

Concluding this article, we would like to point out the importance of small events in understanding the energy release process. Although, as demonstrated here, such phenomena may turn out to be rather complex under high spatial and temporal resolution, they are still much simpler to study than large events and thus allow more direct testing of theoretical models. They thus represent a very interesting class of phenomena for further study.

\begin{acknowledgements}
The authors thank the NRH staff for providing the original visibility data; particular thanks are due to Claude Mercier for his assistance with the self calibration of the NRH visibilities. The authors gratefully acknowledge use of data from the SDO, Solar Monitor, Callisto and GOES/SXI data bases
.The research of A.N. and S.P. has been partly co-financed by the European Union (European Social Fund -ESF) and Greek national funds through the Operational Program ``Education and Lifelong Learning" of the National Strategic Reference Framework (NSRF) -Research Funding Program: ``Thales. Investing in knowledge society through the European Social Fund". S.P. acknowledges support from an FP7 Marie Curie Grant (FP7-PEOPLE-2010-RG/268288).
\end{acknowledgements}

{}

\end{document}